\documentclass[pof,twocolumn,amsmath,amssymb,reprint,nofootinbib]{revtex4-2}
\usepackage{epsfig}                                                                 
\usepackage{dcolumn}                                                                
\usepackage{bm}                                                                     
\usepackage{xcolor}

\PassOptionsToPackage{hyphens}{url}
\usepackage[pdfstartview=FitH,bookmarksopen=true,bookmarksopenlevel=2]{hyperref}    
\usepackage{url}

\begin{document}
\title[Multispecies BGK models and the Onsager reciprocal relations]{Multispecies Bhatnagar--Gross--Krook models and the Onsager reciprocal relations}
\author{E. S. Benilov}
 \email[Email address: ]{Eugene.Benilov@ul.ie}
 \homepage[\newline Homepage: ]{https://eugene.benilov.com/}
 \affiliation{Department of Mathematics and Statistics, University of Limerick, Limerick V94~T9PX, Ireland}

\begin{abstract}
It is shown that most of the existing versions of the Bhatnagar--Gross--Krook
model -- those whose coefficient are independent of the molecular velocity --
do not satisfy the Onsager relations. This circumstance poses a problem when
calibrating these models, making their transport properties match those of a
specific fluid.

\end{abstract}
\maketitle

\section{Introduction\label{sec1}}

In their seminal 1954 paper \cite{BhatnagarGrossKrook54}, Bhatnagar, Gross and
Krook (BGK) proposed a phenomenological model describing kinetic processes in
a pure gas, and two years later, Gross and Krook extended this result to gas
mixtures \cite{GrossKrook56}. Even though neither of these models follows from
the first principles, they are believed to provide a qualitatively correct
approximation of the Boltzmann kinetic equation, and a lot of work has been
done to generalize and extend the BGK approach. In application to mixtures,
the effort has mostly gone into making the BGK model more adaptable, so that
it would be able to describe a wide range of real fluids (e.g., Refs.
\cite{Hamel65,Hamel66,Greene73,Cercignani88,SofoneaSekerka01,Asinari08,HaackHauckMurillo17a,HaackHauckMurillo17b,BobylevBisiGroppiSpigaPotapenko18,TodorovaSteijl19,BoscarinoChoGroppiRusso21,HaackHauckKlingenbergPirnerWarnecke21,PirnerWarnecke22,HaackHauckKlingenbergPirnerWarnecke23,BrullGuillonThieullen24}%
).

Note, however, that the multispecies BGK model has never been tested for
compliance with the Onsager reciprocal relations, which impose certain
constraints on the transport coefficients. Models derived from the first
principles satisfy them automatically, whereas phenomenological models may or
may not do so. An example of a non-compliant model can be viewed in the Enskog
theory of dense fluids \cite{Enskog22}, and an example of a compliant one, in
the so-called modified Enskog theory
\cite{VanbeijerenErnst73a,VanbeijerenErnst73b}. The noncompliance with the
Onsager relations casts doubt on the model's physical relevance, and it is no
coincidence that the modified Enskog theory has eventually been shown to
follow from the first principles for a fluid of hard spheres
\cite{ResiboisLeener77,DorfmanBeijeren77}.

As demonstrated in the present paper, the most common version of the BGK model
(which includes the original result of Gross and Krook \cite{GrossKrook56} as
a particular case) does not comply with the Onsager reciprocal relations.
According to one of those, the coefficient of the temperature gradient in the
mass flux should be inter-linked in a certain way with the coefficient of the
density gradient in the heat flux. According to the BGK model, however, the
former is zero, whereas the latter is proportional to the coefficient of the
density gradient in the mass flux -- hence, can\emph{not} be zero. Not only
does this undermine the physical relevance of the model, this also makes the
BGK model impossible to calibrate -- i.e., choose the values of the parameters
involved to ensure that the transport properties of the fluid under
consideration are described correctly.

In Sec. \ref{sec2} of this paper, one the most general BGK-type models will be
formulated, and in Sec. \ref{sec3}, it will be shown to not comply with the
Onsager relations. Other BGK-type models are briefly discussed in Sec.
\ref{sec4}. For simplicity, only binary (two-species) mixtures will be
considered, but the resulting conclusions apply to the general case as well.

\section{Formulation: the BGK model for binary mixtures\label{sec2}}

Consider a mixture of two monatomic gases, described by the distribution
functions $f_{i}(t,\mathbf{r},\mathbf{v})$, where $i$ is the species number,
$t$ is the time, $\mathbf{r}$ is the position vector, and $\mathbf{v}$, the
molecular velocity. The macroscopic number density $n_{i}$, velocity
$\mathbf{V}_{i}$, and temperature $T_{i}$ of the $i$-th species are given by%
\begin{equation}
n_{i}=%
{\displaystyle\int}
f_{i}\mathrm{d}^{3}\mathbf{v}, \label{2.1}%
\end{equation}%
\begin{equation}
n_{i}\mathbf{V}_{i}=%
{\displaystyle\int}
\mathbf{v}f_{i}\mathrm{d}^{3}\mathbf{v}, \label{2.2}%
\end{equation}%
\begin{equation}
3n_{i}T_{i}=%
{\displaystyle\int}
m_{i}\left\vert \mathbf{v}-\mathbf{V}_{i}\right\vert ^{2}f_{i}\mathrm{d}%
^{3}\mathbf{v}, \label{2.3}%
\end{equation}
where $m_{i}$ is the molecular mass, and $T_{i}$ is measured in energy units
(so that the Boltzmann constant equals unity).

The most general form of the multispecies BGK model (e.g.,
\cite{KlingenbergPirnerPuppo17,PirnerWarnecke22}) consists in%
\begin{align}
\frac{\partial f_{1}}{\partial t}+\mathbf{v}\cdot\mathbf{\nabla}f_{1}  &
=\nu_{11}\left(  M_{1}-f_{1}\right)  +\nu_{12}\left(  M_{12}-f_{1}\right)
,\label{2.4}\\
\frac{\partial f_{2}}{\partial t}+\mathbf{v}\cdot\mathbf{\nabla}f_{2}  &
=\nu_{22}\left(  M_{2}-f_{2}\right)  +\nu_{21}\left(  M_{21}-f_{2}\right)  ,
\label{2.5}%
\end{align}
where $\nu_{ij}$ are the frequencies of collisions between the molecules of
$i$-th and $j$-th species,%
\begin{equation}
M_{i}=n_{i}\left(  \frac{m_{i}}{2\pi T_{i}}\right)  ^{3/2}\exp\left(
-\frac{m_{i}\left\vert \mathbf{v}-\mathbf{V}_{i}\right\vert ^{2}}{2T_{i}%
}\right)  , \label{2.6}%
\end{equation}%
\begin{align}
M_{12}  &  =\left(  \frac{m_{1}}{2\pi T_{12}}\right)  ^{3/2}n_{1}\exp\left(
-\frac{m_{1}\left\vert \mathbf{v}-\mathbf{V}_{12}\right\vert ^{2}}{2T_{12}%
}\right)  ,\label{2.7}\\
M_{21}  &  =\left(  \frac{m_{2}}{2\pi T_{21}}\right)  ^{3/2}n_{2}\exp\left(
-\frac{m_{2}\left\vert \mathbf{v}-\mathbf{V}_{21}\right\vert ^{2}}{2T_{21}%
}\right)  , \label{2.8}%
\end{align}
are various Maxwellian distributions, and%
\begin{align}
\mathbf{V}_{12}  &  =\mathbf{V}_{1}+\beta_{1}\left(  \mathbf{V}_{2}%
-\mathbf{V}_{1}\right)  ,\label{2.9}\\
\mathbf{V}_{21}  &  =\mathbf{V}_{2}+\beta_{2}\left(  \mathbf{V}_{1}%
-\mathbf{V}_{2}\right)  , \label{2.10}%
\end{align}%
\begin{align}
T_{12}  &  =T_{1}+\alpha_{1}\left(  T_{2}-T_{1}\right)  +\gamma_{1}\left\vert
\mathbf{V}_{1}-\mathbf{V}_{2}\right\vert ^{2},\label{2.11}\\
T_{21}  &  =T_{2}+\alpha_{2}\left(  T_{1}-T_{2}\right)  +\gamma_{2}\left\vert
\mathbf{V}_{2}-\mathbf{V}_{1}\right\vert ^{2}. \label{2.12}%
\end{align}
Note that the parameters $\nu_{ij}$, $\alpha_{i}$, $\beta_{i}$, and
$\gamma_{i}$ may depend on the macroscopic characteristics $n_{1}$, $n_{2}$,
$\mathbf{V}_{1}$, $\mathbf{V}_{2}$, etc. -- hence, may vary with $t$ and
$\mathbf{r}$, but not with $\mathbf{v}$. Various particular cases of model
(\ref{2.1})--(\ref{2.12}) have been examined in Refs.
\cite{GrossKrook56,Hamel65,Greene73,Cercignani88,GarzoSantosBrey89,SofoneaSekerka01,Asinari08,BobylevBisiGroppiSpigaPotapenko18}%
.

Eqs. (\ref{2.1})--(\ref{2.12}) form a closed set for $f_{1}(t,\mathbf{r}%
,\mathbf{v})$ and $f_{2}(t,\mathbf{r},\mathbf{v})$. One can readily show that
they conserve mass -- i.e., satisfy%
\begin{equation}
\frac{\partial n_{1}}{\partial t}+\mathbf{\nabla}\cdot\left(  n_{1}%
\mathbf{V}_{1}\right)  =0,\qquad\frac{\partial n_{2}}{\partial t}%
+\mathbf{\nabla}\cdot\left(  n_{2}\mathbf{V}_{2}\right)  =0. \label{2.13}%
\end{equation}
As for the momentum and energy, Eqs. (\ref{2.1})--(\ref{2.12}) do not conserve
them automatically, but only subject to the following constraints:%
\begin{equation}
\alpha_{1}=\frac{\alpha}{\nu_{12}},\qquad\alpha_{2}=\frac{\alpha}{\nu_{21}},
\label{2.14}%
\end{equation}%
\begin{equation}
\beta_{1}=\frac{\beta}{\nu_{12}m_{1}},\qquad\beta_{2}=\frac{\beta}{\nu
_{21}m_{2}}, \label{2.15}%
\end{equation}%
\begin{align}
\gamma_{1}  &  =\frac{1}{3\nu_{12}}\left(  \beta-\frac{\beta^{2}}{\nu
_{12}m_{1}}+3\gamma\right)  ,\label{2.16}\\
\gamma_{2}  &  =\frac{1}{3\nu_{21}}\left(  \beta-\frac{\beta^{2}}{\nu
_{21}m_{2}}-3\gamma\right)  , \label{2.17}%
\end{align}
where the coefficients $\alpha$, $\beta$, and $\gamma$ may depend on
$\mathbf{r}$ and $t$. The above constraints are equivalent to those derived in
Refs. \cite{KlingenbergPirnerPuppo17,PirnerWarnecke22}, albeit presented in a
different form.

Given constraints (\ref{2.14})--(\ref{2.17}), Eqs. (\ref{2.1})--(\ref{2.12})
imply that%
\begin{multline}
\frac{\partial\left(  m_{1}n_{1}\mathbf{V}_{1}+m_{2}n_{2}\mathbf{V}%
_{2}\right)  }{\partial t}\\
+\mathbf{\nabla}\cdot%
{\displaystyle\int}
\mathbf{v}\otimes\mathbf{v}\left(  m_{1}f_{1}+m_{2}f_{2}\right)
\mathrm{d}^{3}\mathbf{v}=0, \label{2.18}%
\end{multline}%
\begin{multline}
\frac{\partial}{\partial t}\left(  \frac{3}{2}n_{1}T_{1}+\frac{m_{1}\left\vert
\mathbf{V}_{1}\right\vert ^{2}}{2}+\frac{3}{2}n_{2}T_{2}+\frac{m_{1}\left\vert
\mathbf{V}_{2}\right\vert ^{2}}{2}\right) \\
+\mathbf{\nabla}\cdot%
{\displaystyle\int}
\frac{\left\vert \mathbf{v}\right\vert ^{2}}{2}\mathbf{v}\left(  m_{1}%
f_{1}+m_{2}f_{2}\right)  \mathrm{d}^{3}\mathbf{v}=0, \label{2.19}%
\end{multline}
which reflect the momentum and energy conservation, respectively.

\section{Transport fluxes under the diffusion approximation\label{sec3}}

\subsection{The standard hydrodynamics\label{sec3.1}}

Within the framework of the Enskog--Chapman approach (e.g., Ref.
\cite{FerzigerKaper72}, chapter 6), the mass and heat fluxes are given by%
\begin{equation}
\mathbf{J}_{i}=-m_{i}n_{i}\left(  \sum_{j}D_{ij}\mathbf{d}_{j}+B_{i}%
\frac{\mathbf{\nabla}T}{T}\right)  ,\label{3.1}%
\end{equation}%
\begin{equation}
\mathbf{Q}=-\kappa\mathbf{\nabla}T-\sum_{i}C_{i}\mathbf{d}_{i},\label{3.2}%
\end{equation}
where%
\begin{multline}
\mathbf{d}_{j}=\mathbf{\nabla}\frac{n_{j}}{n_{1}+n_{2}}\\
+\left(  \frac{n_{j}}{n_{1}+n_{2}}-\frac{\rho_{j}}{m_{1}n_{1}+m_{2}n_{2}%
}\right)  \frac{\mathbf{\nabla}p}{p},\label{3.3}%
\end{multline}
is the \textquotedblleft diffusion driving force\textquotedblright\ of the
$i$-th species, and the pressure is%
\begin{equation}
p=\left(  n_{1}+n_{2}\right)  T.\label{3.4}%
\end{equation}
$D_{ij}$, $B_{i}$, $C_{i}$, and $\kappa$ are the transport coefficients:
$D_{ij}$ is the diffusivity, $B_{i}$ is the thermodiffusivity (it describes
the Soret effect, i.e., the mass flux due to a temperature gradient), $C_{i}$
describes the Dufour effect (i.e., heat flux due to a concentration gradient),
and $\kappa$ is the thermal conductivity.

Most importantly, $C_{i}$ is linked to $B_{i}$ via one of the Onsager
reciprocal relations,%
\[
C_{i}=pB_{i}.
\]
and the other Onsager relation requires that the diffusivity matrix be
symmetric,%
\[
D_{ij}=D_{ji}.
\]
In addition to the above relations, the coefficients $D_{ij}$ and $B_{i}$
should also satisfy%
\[
\sum_{i}D_{ij}=0,\qquad\sum_{i}B_{i}=0
\]
(see Ref. \cite{FerzigerKaper72}). Thus, for a binary mixture, one can express
$D_{ij}$, $B_{i}$, and $C_{i}$ through only two coefficients -- say, $D$ and
$B$ -- so that%
\begin{equation}
D_{11}=\frac{\rho_{2}}{\rho_{1}}D,\qquad D_{22}=\frac{\rho_{1}}{\rho_{2}%
}D,\label{3.5}%
\end{equation}%
\begin{equation}
D_{21}=D_{12}=-D,\label{3.6}%
\end{equation}%
\begin{equation}
B_{1}=\frac{B}{m_{1}n_{1}},\qquad B_{2}=-\frac{B}{m_{2}n_{2}},\label{3.7}%
\end{equation}%
\begin{equation}
C_{1}=\frac{\left(  n_{1}+n_{2}\right)  T}{m_{2}n_{2}}B,\qquad C_{2}%
=-\frac{\left(  n_{1}+n_{2}\right)  T}{m_{1}n_{1}}B.\label{3.8}%
\end{equation}
In this paper, expressions (\ref{3.1})--(\ref{3.8}) will be used under the
diffusion approximation -- which includes the isobaricity assumption (more
details given later) -- i.e., $p\approx\operatorname{const}$. Thus,
expressions (\ref{3.1})--(\ref{3.8}) yield%
\begin{multline}
\mathbf{J}_{1}\approx-D\frac{\left(  m_{1}n_{1}+m_{2}n_{2}\right)  \left(
n_{2}\mathbf{\nabla}n_{1}-n_{1}\mathbf{\nabla}n_{2}\right)  }{\left(
n_{1}+n_{2}\right)  ^{2}}\\
+B\frac{\mathbf{\nabla}n_{1}+\mathbf{\nabla}n_{2}}{n_{1}+n_{2}},\label{3.9}%
\end{multline}%
\begin{multline}
\mathbf{J}_{2}\approx-D\frac{\left(  m_{1}n_{1}+m_{2}n_{2}\right)  \left(
n_{1}\mathbf{\nabla}n_{2}-n_{2}\mathbf{\nabla}n_{1}\right)  }{\left(
n_{1}+n_{2}\right)  ^{2}}\\
-B\frac{\mathbf{\nabla}n_{1}+\mathbf{\nabla}n_{2}}{n_{1}+n_{2}},\label{3.10}%
\end{multline}%
\begin{equation}
\mathbf{Q}\approx-B\frac{T\left(  n_{1}\mathbf{\nabla}n_{2}-n_{2}%
\mathbf{\nabla}n_{1}\right)  }{m_{1}n_{1}m_{2}n_{2}\left(  n_{1}+n_{2}\right)
^{2}}-\kappa\frac{\mathbf{\nabla}n_{1}+\mathbf{\nabla}n_{2}}{\left(
n_{1}+n_{2}\right)  ^{2}}.\label{3.11}%
\end{equation}
In the next subsection, these expressions will be compared to their BGK
counterparts. This is, generally, how the latter could be calibrated, so that
its coefficients are related to the measured values of $D$, $B$, and $\kappa$
of the gas mixture under consideration.

\subsection{The multispecies BGK model\label{sec3.2}}

To derive the hydrodynamic approximation of a kinetic model, one should assume
that the spatial scale of the solution exceeds the length $l$ of the free
path, and the solution's temporal scale exceeds $l/v$ where $v$ is the mean
velocity. Mathematically, these assumptions amount to `stretching' the
coordinates and time -- i.e., replacing%
\begin{equation}
\frac{\partial}{\partial t}\rightarrow\varepsilon\frac{\partial}{\partial
t},\qquad\mathbf{\nabla}\rightarrow\varepsilon\mathbf{\nabla}, \label{3.12}%
\end{equation}
where $\varepsilon$ is a small parameter. One should then assume that the
distribution function is nearly Maxwellian, with the velocity $\mathbf{V}$ and
temperature $T$ being the same for all the species, i.e.,
\begin{multline}
f_{i}=n_{i}\left(  \frac{m_{i}}{2\pi T}\right)  ^{3/2}\exp\left(  -\frac
{m_{i}\left\vert \mathbf{v}-\mathbf{V}\right\vert ^{2}}{2T}\right) \\
+\varepsilon f_{i}^{(1)}+\mathcal{O}(\varepsilon^{2}). \label{3.13}%
\end{multline}
In application to a BGK-type model, the hydrodynamic approximation was
considered in Ref. \cite{HaackHauckMurillo17a}, who also let%
\begin{align}
\mathbf{V}_{i}  &  =\mathbf{V}+\varepsilon\mathbf{V}_{i}^{(1)}+\mathcal{O}%
(\varepsilon^{2}),\label{3.14}\\
T_{i}  &  =T+\varepsilon T_{i}^{(1)}+\mathcal{O}(\varepsilon^{2}),
\label{3.15}%
\end{align}
while leaving $n_{i}$ nonexpanded. Substituting (\ref{3.13})--(\ref{3.15})
into the rescaled versions of Eqs. (\ref{2.4})--(\ref{2.5}), one can find
$f_{i}^{(1)}$ and then use Eqs. (\ref{2.2})--(\ref{2.3}) to find
$\mathbf{V}_{i}^{(1)}$ and $T_{i}^{(1)}$, while Eq. (\ref{2.1}) for $n_{i}$
does not seem to be needed. Such a non-straightforward procedure was chosen in
Ref. \cite{HaackHauckMurillo17a} because of the highly-nonlinear structure of
the BGK\ model, making the straightforward calculation of higher-order
corrections, such as the transport fluxes, cumbersome.

In the present paper, a slightly different approach is employed, where the
transport coefficients are calculated under the \emph{diffusion} approximation
instead of the \emph{hydrodynamic} one. The difference between the two
approximations is two-fold. Firstly, the diffusion flow is slow, so that
scaling (\ref{3.12}) should be replaced with%
\begin{equation}
\frac{\partial}{\partial t}\rightarrow\varepsilon^{2}\frac{\partial}{\partial
t},\qquad\mathbf{\nabla}\rightarrow\varepsilon\mathbf{\nabla.} \label{3.16}%
\end{equation}
Secondly, the diffusion flow is weak -- so that expansions (\ref{3.13}%
)--(\ref{3.14}) should be replaced with%
\begin{multline}
f_{i}=n_{i}^{(0)}\left(  \frac{m_{i}}{2\pi T}\right)  ^{3/2}\exp\left(
-\frac{m_{i}\left\vert \mathbf{v}\right\vert ^{2}}{2T}\right) \\
+\varepsilon f_{i}^{(1)}+\mathcal{O}(\varepsilon^{2}). \label{3.17}%
\end{multline}%
\begin{equation}
\mathbf{V}_{i}=\varepsilon\mathbf{V}_{i}^{(1)}+\mathcal{O}(\varepsilon
^{2}),\qquad T_{i}=T+\varepsilon T_{i}^{(1)}+\mathcal{O}(\varepsilon^{2}),
\label{3.18}%
\end{equation}
and the density should also be expanded,
\begin{equation}
n_{i}=n_{i}^{(0)}+\varepsilon n_{i}^{(1)}+\mathcal{O}(\varepsilon^{2}).
\label{3.19}%
\end{equation}
Under such an approximation, the transport fluxes emerge from the
\emph{leading} order of the expansion.

Having rescaled the BGK equations (\ref{2.4})--(\ref{2.5}) according to
(\ref{3.16}), one should substitute into them expansions (\ref{3.17}%
)--(\ref{3.19}). $\mathbf{V}_{ij}$ and $T_{ij}$ should also be expanded
[similarly to how $\mathbf{V}_{i}$ and $T_{i}$ are expanded in (\ref{3.18})],
as well as all the coefficients,\begin{widetext}%
\[
\nu_{ij}=\nu_{ij}^{(0)}+\mathcal{O}(\varepsilon),\qquad\alpha_{i}=\alpha
_{i}^{(0)}+\mathcal{O}(\varepsilon),\qquad\beta_{i}=\beta_{i}^{(0)}%
+\mathcal{O}(\varepsilon),\qquad\gamma_{i}=\gamma_{i}^{(0)}+\mathcal{O}%
(\varepsilon).
\]
Eqs. (\ref{2.4})--(\ref{2.5}) are linear algebraic equations, and one can
readily deduce that%
\begin{multline*}
f_{1}^{(1)}=\left[  \frac{n_{1}^{(1)}}{n_{1}^{(0)}}+\frac{m_{1}\left\vert
\mathbf{v}\right\vert ^{2}-3T}{2T}\frac{\nu_{11}^{(0)}T_{1}^{(1)}+\nu
_{12}^{(0)}T_{12}^{(1)}}{\left(  \nu_{11}^{(0)}+\nu_{12}^{(0)}\right)
T}+m_{1}\mathbf{v}\cdot\frac{\nu_{11}^{(0)}\mathbf{V}_{1}^{(1)}+\nu_{12}%
^{(0)}\mathbf{V}_{12}^{(1)}}{\left(  \nu_{11}^{(0)}+\nu_{12}^{(0)}\right)
T}\right]  M_{1}^{(0)}\\
-\frac{\mathbf{v}}{\nu_{11}^{(0)}+\nu_{12}^{(0)}}\cdot\left(  \frac
{\mathbf{\nabla}n_{1}^{(0)}}{n_{1}^{(0)}}+\frac{m_{1}\left\vert \mathbf{v}%
\right\vert ^{2}-3T}{2T}\frac{\mathbf{\nabla}T}{T}\right)  M_{1}^{(0)},
\end{multline*}%
\begin{multline*}
f_{2}^{(1)}=\left[  \frac{n_{2}^{(1)}}{n_{2}^{(0)}}+\frac{m_{2}\left\vert
\mathbf{v}\right\vert ^{2}-3T}{2T}\frac{\nu_{22}^{(0)}T_{2}^{(1)}+\nu
_{21}^{(0)}T_{21}^{(1)}}{\left(  \nu_{22}^{(0)}+\nu_{21}^{(0)}\right)
T}+m_{2}\mathbf{v}\cdot\frac{\nu_{22}^{(0)}\mathbf{V}_{2}^{(1)}+\nu_{21}%
^{(0)}\mathbf{V}_{21}^{(1)}}{\left(  \nu_{22}^{(0)}+\nu_{21}^{(0)}\right)
T}\right]  M_{2}^{(0)}\\
-\frac{\mathbf{v}}{\nu_{22}^{(0)}+\nu_{21}^{(0)}}\cdot\left(  \frac
{\mathbf{\nabla}n_{2}^{(0)}}{n_{2}^{(0)}}+\frac{m_{1}\left\vert \mathbf{v}%
\right\vert ^{2}-3T}{2T}\frac{\mathbf{\nabla}T}{T}\right)  M_{2}^{(0)},
\end{multline*}
\end{widetext}where%
\[
M_{i}^{(0)}=n_{i}^{(0)}\left(  \frac{m_{i}}{2\pi T}\right)  ^{3/2}\exp\left(
-\frac{m_{i}\left\vert \mathbf{v}\right\vert ^{2}}{2T}\right)  .
\]
Substituting these expressions into the leading order of Eqs. (\ref{2.1}%
)--(\ref{2.3}), one can verify that Eq. (\ref{2.1}) is satisfied identically,
and Eqs. (\ref{2.2})--(\ref{2.3}), (\ref{2.9})--(\ref{2.12}) yield%
\begin{equation}
\mathbf{\nabla}\left(  n_{1}^{(0)}T^{(0)}+n_{2}^{(0)}T^{(0)}\right)  =0,
\label{3.20}%
\end{equation}%
\begin{align}
\mathbf{V}_{1}^{(1)}  &  =\mathbf{V}^{(1)}-\frac{\mathbf{\nabla}\left(
n_{1}^{(0)}T^{(0)}\right)  }{2\beta^{(0)}n_{1}^{(0)}n_{2}^{(0)}}%
,\label{3.21}\\
\mathbf{V}_{2}^{(1)}  &  =\mathbf{V}^{(1)}-\frac{\mathbf{\nabla}\left(
n_{2}^{(0)}T^{(0)}\right)  }{2\beta^{(0)}n_{1}^{(0)}n_{2}^{(0)}}, \label{3.22}%
\end{align}%
\[
T_{12}^{(1)}=T_{1}^{(1)}=T_{21}^{(1)}=T_{2}^{(1)}=T^{(1)},
\]
where $\mathbf{V}^{(1)}(\mathbf{r},t)$ and $T^{(1)}(\mathbf{r},t)$ are
undetermined functions (neither will appear in the final expressions for the
fluxes). Note that $\beta_{1}$ and $\beta_{2}$ which appear in the original
set have been expressed through $\beta$ using (\ref{2.15}).

Physically, Eq. (\ref{3.20}) reflects the isobaric nature of diffusion and
heat conduction (the same result follows from the standard hydrodynamic
equations or any other model). Introducing the leading-order pressure
$p^{(0)}$ (which may depend only on $t$), one can rewrite (\ref{3.20}) in the
form%
\begin{equation}
T^{(0)}=\frac{p^{(0)}}{n_{1}^{(0)}+n_{2}^{(0)}}.\label{3.24}%
\end{equation}
Next, introduce the mass-averaged velocity,%
\[
\mathbf{\bar{V}}=\frac{m_{1}n_{1}\mathbf{V}_{1}+m_{2}n_{2}\mathbf{V}_{2}%
}{m_{1}n_{1}+m_{2}n_{2}},
\]
and the diffusion flux of the $i$-th species,%
\[
\mathbf{J}_{i}=m_{i}n_{i}\left(  \mathbf{V}_{i}-\mathbf{\bar{V}}\right)  .
\]
Using (\ref{3.21})--(\ref{3.22}), one can calculate $\mathbf{J}_{i}$ and then
use (\ref{3.24}) to eventually obtain\begin{widetext}%
\[
\mathbf{J}_{1}=-\varepsilon\frac{T^{(0)}m_{1}m_{2}\left(  n_{2}^{(0)}%
\mathbf{\nabla}n_{1}^{(0)}-n_{1}^{(0)}\mathbf{\nabla}n_{2}^{(0)}\right)
}{\beta^{(0)}\left(  m_{1}n_{1}^{(0)}+m_{2}n_{2}^{(0)}\right)  \left(
n_{1}^{(0)}+n_{2}^{(0)}\right)  }+\mathcal{O}(\varepsilon^{2}),\qquad
\mathbf{J}_{2}=-\varepsilon\frac{T^{(0)}m_{1}m_{2}\left(  n_{1}^{(0)}%
\mathbf{\nabla}n_{2}^{(0)}-n_{2}^{(0)}\mathbf{\nabla}n_{1}^{(0)}\right)
}{2\beta^{(0)}\left(  m_{1}n_{1}^{(0)}+m_{2}n_{2}^{(0)}\right)  \left(
n_{1}^{(0)}+n_{2}^{(0)}\right)  }+\mathcal{O}(\varepsilon^{2}).
\]
Comparing these expressions to their `correct' counterparts (\ref{3.9}%
)--(\ref{3.10}), one can see that the two results can be reconciled by
choosing a certain value of $\beta$ only if $B=0$ (no thermodiffusivity). One
might think that the BGK model may still work for mixtures whose
thermodiffusivity is indeed small -- e.g., that of water vapor and air (see
the estimates in Refs. \cite{LidonPerrotStroock21,Benilov23a}) -- but,
unfortunately, a further problem arises even in this case. To illustrate it,
consider the heat flux,%
\[
\mathbf{Q}=\int\frac{\left\vert \mathbf{v}\right\vert ^{2}}{2}\mathbf{v}%
\left(  m_{1}f_{1}+m_{2}f_{2}\right)  \mathrm{d}^{3}\mathbf{v}-5\left(
n_{1}T_{1}+n_{2}T_{2}\right)  \mathbf{\bar{V}},
\]
which, to leading order, is%
\begin{multline}
\mathbf{Q}=-\varepsilon\frac{5T^{(0)2}\left(  m_{2}-m_{1}\right)  \left(
n_{2}^{(0)}\mathbf{\nabla}n_{1}^{(0)}-n_{1}^{(0)}\mathbf{\nabla}n_{2}%
^{(0)}\right)  }{\beta^{(0)}\left(  m_{1}n_{1}^{(0)}+m_{2}n_{2}^{(0)}\right)
\left(  n_{1}^{(0)}+n_{2}^{(0)}\right)  }\\
+\varepsilon\left[  \frac{n_{1}^{(0)}}{m_{1}\left(  \nu_{11}^{(0)}+\nu
_{12}^{(0)}\right)  }+\frac{n_{2}^{(0)}}{m_{2}\left(  \nu_{22}^{(0)}+\nu
_{21}^{(0)}\right)  }\right]  \dfrac{5T^{(0)2}\left(  \mathbf{\nabla}%
n_{1}^{(0)}+\mathbf{\nabla}n_{2}^{(0)}\right)  }{n_{1}^{(0)}+n_{2}^{(0)}%
}+\mathcal{O}(\varepsilon^{2}).\label{3.25}%
\end{multline}
\end{widetext}Comparing this expression to its `correct' counterpart
(\ref{3.11}) with $B=0$, one can see that the two results coincide only in the
limit $\beta\rightarrow\infty$ which makes the \emph{whole} diffusive flux
equal zero, not only its thermodiffusive part.

One way or another, no such value of the parameter $\beta$ exists that makes
the BGK fluxes satisfy the Onsager reciprocal relation -- neither for the
general case nor for a fluid with zero thermodiffusivity.

\section{Concluding remarks\label{sec4}}

It should be emphasized that the results of the present paper apply to some,
but not all, of the existing BGK-type models. Apart from the model examined
above, they apply to that of Refs.
\cite{HaackHauckMurillo17a,HaackHauckMurillo17b}, which consists of Eqs.
(\ref{2.1})--(\ref{2.10}), but with Eqs. (\ref{2.11})--(\ref{2.12}) replaced
with%
\begin{align*}
T_{12} &  =T_{1}+\alpha_{1}\left(  T_{2}-T_{1}\right)  +\gamma_{1}\left(
\left\vert \mathbf{V}_{1}\right\vert ^{2}-\left\vert \mathbf{V}_{2}\right\vert
^{2}\right)  ,\\
T_{21} &  =T_{2}+\alpha_{2}\left(  T_{1}-T_{2}\right)  +\gamma_{2}\left(
\left\vert \mathbf{V}_{2}\right\vert ^{2}-\left\vert \mathbf{V}_{1}\right\vert
^{2}\right)  .
\end{align*}
Even though these expressions differ from their counterparts examined here,
the expressions for $\mathbf{V}_{12}$ and $\mathbf{V}_{21}$ are still the
same, and this is enough for noncompliance with the Onsager relations. As for
models where the collision frequencies $\nu_{ij}$ depend on the molecular
velocity (e.g.,
\cite{HaackHauckKlingenbergPirnerWarnecke21,HaackHauckKlingenbergPirnerWarnecke23}%
), those need to be tested separately. The present results do not cover them.

Note also that, even though the models examined in this paper do not formally
satisfy the Onsager relations, they satisfy them \emph{asymptotically} in the limit
\begin{equation}
\frac{m_{1}}{m_{2}}\rightarrow0.\label{3.26}%
\end{equation}
To understand why, observe that the ratio of the first to second terms of heat
flux (\ref{3.25}) is proportional to $m_{1}/m_{2}$ -- hence, condition
(\ref{3.26}) allows one to neglect the first term. After that, expression
(\ref{3.25}) matches the standard heat flux expression (\ref{3.2}) with
$C_{i}=0$, and so the corresponding Onsager relation holds. Note also that
asymptotic limit (\ref{3.26}) is important physically, as it describes ionized
plasma (where the mass of electrons is indeed much smaller than that of ions).
The asymptotic compliance with the Onsager relations occurs also in the limit
$m_{1}/m_{2}\rightarrow1$, in which case the first term in expression
(\ref{3.25}) vanishes.

One should not assume, however, that BGK-type models cannot satisfy the Onsager relations exactly. The model proposed in Ref.
\cite{GarzoSantosBrey89}, for example, does have the correct transport
properties -- and also satisfies the so-called indifferentiability principle
(i.e., if the molecules of the species have identical mechanical parameters,
the distribution function of the mixture satisfies the single-species BGK
equation). Unfortunately, this model does not seem to comply with the H
theorem, as pointed out in Ref. \cite{HaackHauckMurillo17a}.

Overall, a `perfect' multispecies kinetic model should satisfy the following requirements:

\begin{enumerate}
\item conservation of mass, momentum, and energy;

\item H theorem;

\item indifferentiability principle;

\item positivity of the temperature and concentration;

\item ability to represent fluids with arbitrary values of the Prandtl and Schmidt numbers, and an arbitrary ratio of the bulk and shear viscosities;

\item Onsager reciprocal relations.
\end{enumerate}

So far, none of the existing BGK-type models has been shown to comply with
\emph{all} of the above requirements (see, for example, the review sections of
Refs. \cite{BoscarinoChoGroppiRusso21,BrullGuillonThieullen24}). This does not
mean, however, that a fully compliant model does not exist in principle, and
so one should hope that such will be developed in the future.

Finally, note that requirements 5--6 of the above list are particularly
important for end users -- i.e., researchers who need a practical tool to work
with applications (like the present author, who looks for a tool to model
evaporation of water into air). These requirements allow one to calibrate the
model, so that its transport properties match those of the fluid under consideration.


\bibliography{}

\end{document}